\documentclass[12pt,epsfig]{article}
\usepackage{graphicx}
\usepackage{amssymb}
\usepackage{amsmath}

\def\permil{\%\raise.10ex\hbox{$_{\scriptstyle 0}$}}

\def\beq{\begin{equation}}
\def\eeq{\end{equation}}
\def\beqn{\begin{eqnarray}}
\def\eeqn{\end{eqnarray}}

\newcommand{\shat}{\ensuremath{\hat{s}}}
\newcommand{\bhat}{\ensuremath{\hat{b}}}
\newcommand{\that}{\ensuremath{\hat{t}}}

\newcommand{\thatmin}{\ensuremath{\hat{t}_{\rm min}}}

\newcommand{\kommentar}[1]{}
\newcommand{\orders}[1]{\ensuremath{\mathcal{O}\left(s^{#1}\right)}}

\newcommand{\order}[1]{\ensuremath{\mathcal{O}\left(#1\right)}}
\newcommand{\fal}{fixed angle limit}
\newcommand{\rgl}{Regge limit}
\newcommand{\ons}{on-shell}
\newcommand{\ofs}{off-shell}

\newcommand{\cosq}{\cos\theta}
\newcommand{\imag}{\Im \text{m}}
\newcommand{\mbf}[1]{\mbox{\boldmath $#1$}}

\newcommand{\bb}{\mbf{b}}

\newcommand{\bq}{\mbf{q}}

\begin{document}
\hfill
\hspace*{\fill}
\begin{minipage}[t]{3cm}
  DESY-05-083\\
 hep-ph/0506047
\end{minipage}
\vspace*{1.cm}
\begin{center}
\begin{Large}
{\bf Off-Shell Scattering Amplitudes for $WW$ Scattering\\
and the Role of the Photon Pole}\\[1cm]
\end{Large}
\begin{large}
\vspace{0.5cm}
J. Bartels and F. Schwennsen\\[1cm] 
II. Institut f\"{u}r Theoretische Physik, Universit\"{a}t Hamburg\\
Luruper Chaussee 149, D-22761 Hamburg, Germany\\[1cm] 
\end{large}
\end{center}
\vskip15.0pt \centerline{\bf Abstract}
\noindent
We derive analytic expressions for high energy $2 \to 2$ off-shell scattering 
amplitudes of weak vector bosons. They are obtained from six fermion 
final states in processes of the type $e^+ e^- \to \bar\nu_e + (WW) + \nu_e 
\to \bar\nu_e + (l\nu)(l\nu) + \nu_e$. As an application we reconsider the unitarity 
bounds on the Higgs mass. Particular attention is given to the 
role of the photon exchange which has not been considered in earlier 
investigations; we find that the photon weakens the bound of the Higgs mass.

\section{Introduction}

The elastic scattering of weak vector bosons plays a central 
role in our understanding of the electroweak sector of the standard model. 
It is known that, 
without any Higgs, these scattering processes have a bad high energy behavior,
i.e. they start to violate, at energies of about $\sim 1$TeV, 
bounds derived from unitarity. The Higgs mechanism cures this defect; 
but if the Higgs mass is larger than about $\sim 1$TeV \cite{Lee:1977eg}, 
the weak sector becomes strongly interacting, and the use of low order 
perturbation theory becomes insufficient. So there is 
little doubt that the high energy behavior of heavy vector boson scattering  
carries much information on the electroweak symmetry breaking.

Although the field has already a rather long history 
\cite{Cornwall:1974km,Dawson:1985gx,Veltman:1990ud,Barger:1995cn,Kuss:1996yv,Denner:1998kq,Gangemi:1998vc}, 
there still remain aspects which deserve further studies. One is the stability of 
the heavy vector bosons in the electroweak sector. Most of the    
existing studies consider the scattering of on-shell vector mesons, i.e.
the heavy gauge boson is treated as a stable particle. 
A more realistic investigation has to include both the production and the 
decay of the bosons which necessarily leads to off-shell bosons, both in the 
initial and in the final state.
Secondly, a full analysis of boson scattering in the electroweak sector 
includes the photon: because of the photon pole in the $t$-channel, 
on-shell vector boson scattering, strictly speaking, is not defined in the 
forward direction. In this paper, we attempt to address some of these issues 
in the framework of more `realistic' off-shell scattering processes.

\begin{figure}
\begin{center}\includegraphics{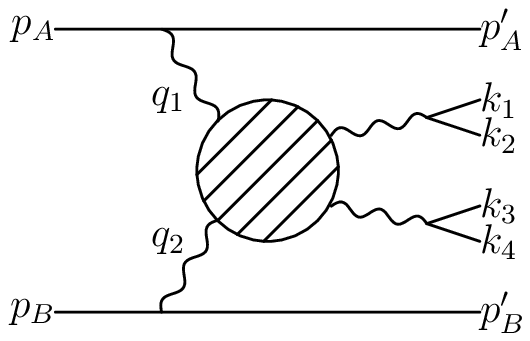}\end{center}
\caption{$e^+e^- \to 6f.$}
\label{figure2nach6}
\end{figure}

One of the possibilities of embedding vector scattering, e.g. the elastic
scattering of two $W$ bosons, 
into a `realistic' process is the six fermion process $e^+e^-$ $\to$ $6f$: 
(Fig.\ref{figure2nach6}). 
In order to isolate `incoming' $W$ bosons for the $WW$ scattering subprocess, 
one has to consider the process $e^+e^- \to 6f$
in the restricted kinematic region where the 
produced $WW$ pair lies in the central region, i.e. in rapidity it is 
well-separated from the fragments of the incoming leptons, the 
$\nu_e$ and the $\bar{\nu}_e$. In this region the $t$-channel $W$'s 
serve as incoming bosons: they are radiated off the incoming 
electron and positron, and their momenta are spacelike and 
off-shell. Also the `outgoing' $W$ bosons are off shell, but, in contrast 
to the `incoming' $W$ bosons,  they have timelike
momenta. In order to simulate $WW$ scattering at high energies, we need 
the subenergy of the $WW$ subprocess to be as large as possible.
A computer-based analysis of this class of six-fermion processes has been given in   
\cite{Dittmaier:2002ap}. In our paper we derive analytic expressions for 
the vector scattering subprocesses which can be used for further theoretical 
investigations. As an example, we 
will study the influence of the Higgs particle on the high energy behavior 
of $WW$ scattering; an aspect of particular interest is the role of the 
photon pole in the unitarity bound.    
      
In section 2 we derive expressions for the process $e^+e^- \to 6f$ 
which are valid in the kinematic 
region described above. In section 3 we discuss a unitarity condition 
for the $WW$ scattering subprocess, and we discuss the role of the photon 
pole. We also comment on the role of the instability of the $W$ boson 
in the unitarity equation.       

\section{The processes $e^+e^-  \to 6f$}
\label{ofs}
We begin with elastic $WW$ - scattering which appears as a subprocess of  
the reaction $e^+e^-  \to \nu_e \;4f\; \bar{\nu}_e.$ Our notations are 
indicated in Fig.1, and  a few diagrams are 
illustrated in Fig.2. In particular, $s=(p_A+p_B)^2$ denotes the squared energy of 
incoming electrons, and $\hat{s} = (q_1 + q_2)^2$ the mass squared of the 
produced $WW$ system.
In order to obtain a meaningfull \ofs\ extension of high energy $WW$ scattering, 
we consider the so-called multi-Regge limit: the squared masses of the 
``incoming'' $W$-momenta, $t_1 = q_1^2$ and $t_2 = q_2^2$, and of the 
``outgoing'' momenta, $M_1^2 =(k_1+k_2)^2\equiv k_{12}^2$ and 
$M_2^2 =(k_3+k_4)^2\equiv k_{34}^2$ are of the order 
\orders{0} =${\mathcal{O}} \left( M_W^2 \right)$.    
Inside the subprocess: $q_1 + q_2 \to k_{12} + k_{34}$ we consider the 
high energy limit $\hat{s} \gg t_1$, $t_2$, $M_1^2$, $M_2^2$, 
$\that = (q_1 - k_{12})^2 \equiv q^2$, where $\that=$ \orders{0}. 
Bounds on the Higgs mass which follow from unitarity constraints, usually,  
are studied in the fixed angle limit, $\hat{s} \sim \hat{t}$. In this paper 
we prefer the Regge limit, $\hat{s} \gg \hat{t}$. In this kinematic region the scattering amplitudes 
for vector - vector scattering take a particularly simple factorized form. 
Further below we will briefly indicate how the known results for on-shell 
scattering, derived in the fixed-angle limit, can be re-derived also in the 
Regge limit. As an important feature of the kinematic limit which we are going to 
investigate, the momentum transfer $\that$ of the subsystem is always 
negative and never reaches 0. The minimal value of $|\hat{t}|$, $-\thatmin$, 
is given by:
\begin{gather}
\thatmin = \frac{(t_1-M_1^2)(t_2-M_2^2)}{\shat}+\order{\shat^{-2}}.
\label{tmin}
\end{gather}
Here the leading term is always less or equal to zero, since  
\begin{gather}   
t_1, t_2 < m_e^2 \quad \text{and} \quad  M_1^2, M_2^2 > m_e^2.
\label{eqthatltzero}
\end{gather}
\begin{figure}
\begin{center}\includegraphics{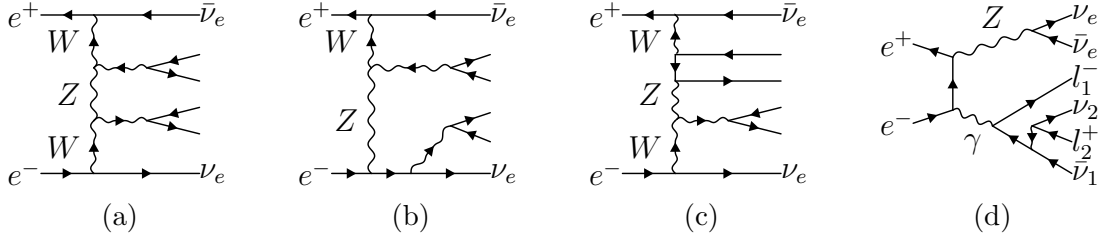}\end{center}
\caption{A few diagrams contributing to the process $e^+e^-  \to \nu_e \;4f\; \bar{\nu}_e.$}
\label{figuresample}
\end{figure}

In the sum of all graphs which contribute to the matrix element we retain only 
those which, in the multi-Regge limit, contribute to the leading power of 
energy. This includes the diagrams with vector bosons in the exchange channel
(Fig.2a,b,c) and eliminates those where fermions are exchanged 
(Fig.2d). Furthermore, we distinguish between `resonant' (Fig.2a,b) 
and `nonresonant' (Fig.2c) graphs: the former ones contain a $W$ pole in the
$M_1$ and in the $M_2$  channel, whereas the latter ones do not. 
Omitting the graphs of type 2b or 2c would lead to incorrect gauge 
dependent results.

The sum of all graphs which contribute to the multi-Regge limit can be written 
in the simple factorizing form:   
\begin{align}
T^{(s)}_{2 \to 4\to 6} =& -2s \Gamma_{l\nu W} \frac{1}{t_1-M_W^2}\nonumber\\
& \Bigg[ 
\frac{\Gamma_{WZ(W\to \nu_1^{\phantom{+}}l_1^+)}(q_1,-q,k_1,k_2)}{M_1^2-M_W^2}\;
\frac{1}{\hat{t}-M_Z^2}\;
\frac{\Gamma_{ZW(W\to l_2^-\bar\nu_2^{\phantom{-}})}(q,q_2,k_3,k_4)}{M_2^2-M_W^2} \nonumber\\ 
& +
\frac{\Gamma_{W \gamma (W\to \nu_1^{\phantom{+}}l_1^+)}(q_1,-q,k_1,k_2)}{M_1^2-M_W^2}\;
\frac{1}{\hat{t}} \;
\frac{\Gamma_{\gamma W(W\to l_2^-\bar\nu_2^{\phantom{-}})}(q,q_2,k_3,k_4)}{M_2^2-M_W^2}
\Bigg]\frac{1}{t_2-M_W^2} \Gamma_{l\nu W}.
\label{eq2n4n6}  
\end{align}

Here $\Gamma_{l\nu W}$ denotes the coupling of the virtual $W$ boson to 
the incident electron or positron.
In the high energy limit the helicity flip 
amplitudes are supressed. Therefore, the coupling contains a helicity 
conserving Kronecker $\delta_{\lambda \lambda'}$ where $\lambda$ ($\lambda'$) 
refer to the incoming (outgoing) fermion.
Since $W$ bosons only couple to left handed fermions, an additional 
$\delta_{\lambda,-}$ appears.
Other vector scattering processes, such as $WW\to ZZ$, $\gamma\gamma\to WW$, 
or $WZ\to WZ$ factorize in the same way. We therefore present the results for 
these couplings in a general form which is applicable also to these processes. 
Because of its different couplings to left and right handed fermions, the 
emission of a virtual $Z$ has a term proportional to 
$\delta_{\lambda,-}$, and another one proportional to $\delta_{\lambda,+}$ 
The emission of a virtual photon, on the other hand, is chiral invariant. 
We obtain: 
\begin{align}
\Gamma_{l\nu W}=& \frac{g_w}{\sqrt{2}} \delta_{\lambda\lambda'}\delta_{\lambda,-}&
\Gamma_{ll Z} =& \frac{g_w}{2c_w} \delta_{\lambda\lambda'} 
\left( (s_w^2-c_w^2)\delta_{\lambda,-}+2s_w^2 \delta_{\lambda,+}\right)&
\Gamma_{ll \gamma} =& g_w s_w \delta_{\lambda\lambda'},
\label{productionvertices}
\end{align}
where $s_w=\sin \Theta_W$ and $c_w=\cos \Theta_W$ are our shorthand 
notations for the sine and cosine of the Weinberg angle, and $g_w$ denotes 
the weak coupling constant which is connected to the Fermi constant $G_F$ 
via $\frac{g_w^2}{8M_W^2}=\frac{G_F}{\sqrt{2}}$. 

Next we turn to the effective production vertices on the rhs of 
(\ref{eq2n4n6}). For the general subprocess 
$V_1V_2\to V_3\to f_a\bar{f}_b$ (Fig.3) it is of the form:
\begin{multline}
\Gamma_{V_1V_2(V_3\to f_a\bar{f}_b )}(p_1,p_2,p_a,p_b) =
\chi\widehat{\Gamma}_{V_1V_2V_3}^{\rho}\bar{u}_{f_a} \gamma_{\rho} d v_{\bar{f}_b}+\big((p_1+p_2)^2-M_{V_3}^2\big)\Gamma'_{V_1V_2(V_3\to f_a\bar{f}_b )}.
\label{eqVVffVertex}
\end{multline}
\begin{figure}[h!]
\begin{center}\includegraphics{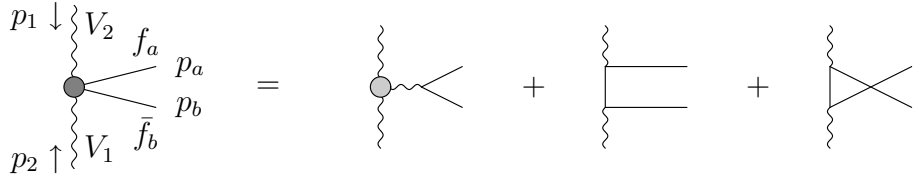}\end{center}
\caption{The effective vertex $\Gamma_{V_1V_2(V_3\to f_a\bar{f}_b )}(p_1,p_2,p_a,p_b)$}\label{figvertex}
\end{figure}

Here $V_1$ and $V_2$ are the upper and the lower `incoming' virtual vector 
bosons with momenta $p_1$, $p_2$ and masses $M_{V_1}$, $M_{V_2}$, resp. 
The part $\widehat{\Gamma}^\rho_{V_1V_2V_3}$ on the rhs of 
eq.(\ref{eqVVffVertex}) contains the `resonant' graphs with a particle 
pole of the produced vector boson, while 
$\Gamma'_{V_1V_2(V_3\to f_a\bar{f}_b )}$ stands for the `nonresonant' ones 
(last two diagrams of figure \ref{figvertex}). The analytic form of 
$\widehat{\Gamma}^\rho_{V_1V_2V_3}$ is ~\cite{BFKL}: 
\begin{equation}
\widehat{\Gamma}^\rho_{V_1V_2V_3} = (p_{1\perp}-p_{2\perp})^\rho-\left(\alpha_1+2\frac{p_1^2-M_{V_1}^2}{\beta_2 s}\right) p_A^\rho+\left(\beta_2+2\frac{p_2^2-M_{V_2}^2}{\alpha_1 s}\right) p_B^\rho. \label{eqgammahat}
\end{equation}
Here we have made use of the Sudakov decomposition of the momenta $p_i=\alpha_i p_A+\beta_i p_B+p_{i\perp}$. 
The coefficients $\chi$ denote the triple vector boson coupling constant:
\begin{align}
\chi(WZ(W\to \nu_1^{\phantom{+}}l_1^+))=& g_w c_w & 
\chi(W\gamma(W\to \nu_1^{\phantom{+}}l_1^+) )=& -g_w s_w.
\label{eqchi}
\end{align}
All other couplings $\chi$ 
can be obtained by using the relations $\chi(V_1V_2V_3)= \chi(V_2V_3V_1)$ and 
$\chi(V_1V_2V_3)= -\chi(V_2V_1V_3)$. The $\gamma$-matrices $d$ describe the coupling of $V_3$ to its decay products:
\begin{equation}
\begin{split}
d\left(W^{+}\to\nu l^{+}\right)=d\left( W^{-}\to l^{-}\bar\nu\right)=&\frac{g_w}{\sqrt{2}}\omega_L\\
d\left(Z\to \nu\bar\nu\right)=& \frac{g_w}{2c_w}\omega_L\\
d\left(Z\to l^{-}l^{+}\right)=&\frac{g_w}{2c_w}\left( (s_w^2-c_w^2)\omega_L+2s_w^2 \omega_R\right)\\
d\left(\gamma\to l^{-}l^{+}\right)=& g_ws_w,
\label{eqkappa3}
\end{split}
\end{equation}
where $\omega_L = \frac{1}{2}(1-\gamma_5)$ 
($\omega_R = \frac{1}{2}(1+\gamma_5)$) project on the 
left handed (right handed) spinors.

For the nonresonant production vertex we find:
\begin{equation} 
\Gamma'_{V_1V_2(V_3\to f_a\bar{f}_b )} = \bar{u}_{f_a}\left(\kappa_1\frac{\underline{p_{\bar{f}_b }}-\underline{p_1}}{(p_{\bar{f}_b}-p_1)^2}+\kappa_2\frac{\underline{p_{f_a}^\star}-\underline{p_1^\star}}{(p_{f_a}-p_1)^2}\right)\omega_L v_{\bar{f}_b}.\label{eqgammaprime}
\end{equation}
In this expression  we have introduced 
the convention $\underline{p}=(p_x+i p_y)(\gamma^1-i \gamma^2)$, 
$\underline{p}^{\star}=(p_x-i p_y)(\gamma^1+i \gamma^2)$. The coefficients $\kappa_{1}$ and $\kappa_{2}$ are given by
\begin{equation}
\begin{split}
\kappa_1\left(WZ\nu l^{+}\right)=\kappa_1\left(ZWl^{-}\bar\nu\right)=& \frac{g_w^2}{2\sqrt{2}c_w}\\
\kappa_2\left(WZ\nu l^{+}\right)=\kappa_2\left(ZWl^{-}\bar\nu\right)=&\frac{g_w^2c_w(2M_W^2-M_Z^2)}{2\sqrt{2}M_W^2}\\
\kappa_1\left(W\gamma\nu l^{+}\right)=\kappa_1\left(\gamma Wl^{-}\bar\nu\right)=& 0\\
\kappa_2\left(W\gamma\nu l^{+}\right)=\kappa_2\left(\gamma Wl^{-}\bar\nu\right)=&-\frac{g_w^2s_w}{\sqrt{2}}\\
\kappa_1\left(W^{(+)}W^{(-)}\nu\bar\nu\right)=\kappa_2\left(W^{(+)}W^{(-)}l^{-}l^{+}\right)=&0 \\
\kappa_2\left(W^{(+)}W^{(-)}\nu\bar\nu\right)=&-g_w^2\frac{2M_W^2-M_Z^2}{2M_W^2}\\
\kappa_1\left(W^{(+)}W^{(-)}(Z\to l^{-}l^{+})\right)=&g_w^2c_w^2\frac{2M_W^2-M_Z^2}{2M_W^2}\\
\kappa_1\left(W^{(+)}W^{(-)}(\gamma\to l^{-}l^{+})\right)=&g_w^2s_w^2,
\end{split}
\label{eqkappa12}
\end{equation}
where the remaining couplings $\kappa_{1}$ and $\kappa_{2}$ can be 
constructed via $\kappa_1(V_1V_2f_a\bar{f}_b)=-\kappa_2(V_2V_1f_a\bar{f}_b)$.

With these building blocks it is easy to construct the scattering amplitudes 
for all \ofs\ vector boson scattering processes in the multi-Regge limit. 
Elastic $WW$ scattering is a subprocesss of the process $e^+e^- \to 
\bar\nu_e (\nu l^+)(l^- \bar \nu) \nu_e$, and the amplitude has already been 
given in 
(\ref{eq2n4n6}). The subprocesses $WW\to ZZ$ and $WW\to \gamma\gamma$ 
are contained in the $2 \to 6$ process 
$e^+e^-$ $\to$ $\bar\nu_e (l_1^-l_1^+)(l_2^-l_2^+) \nu_e$. The corresponding 
$2 \to 6$ scattering amplitudes are:  

\begin{subequations}
\begin{multline}
-2s\Gamma_{l\nu W}\frac{1}{t_1-M_W^2}\\
\left[\frac{\Gamma_{WW(Z\to l_1^-l_1^+)}(q_1,-q,k_1,k_2)}{M_1^2-M_Z^2}\frac{1}{\that-M_W^2}\frac{\Gamma_{WW(Z\to l_2^-l_2^+)}(q,q_2,k_3,k_4)}{M_2^2-M_Z^2}\right]
\frac{1}{t_2-M_W^2}\Gamma_{l\nu W},
\label{example2}
\end{multline}
\begin{multline}
 -2s\Gamma_{l\nu W}\frac{1}{t_1-M_W^2}\\
\left[\frac{\Gamma_{WW(\gamma\to l_1^-l_1^+)}(q_1,-q,k_1,k_2)}{M_1^2}\frac{1}{\that-M_W^2}\frac{\Gamma_{WW(\gamma\to l_2^-l_2^+)}(q,q_2,k_3,k_4)}{M_2^2}\right]
\frac{1}{t_2-M_W^2}\Gamma_{l\nu W}.
\end{multline}
\end{subequations}
Similarly, the subprocesses $ZZ\to WW$ and $\gamma\gamma\to WW$ are part of 
the $2\to 6$ process 
$e^+e^-$ $\to$ $e^+ (\nu_1^{\vphantom{+}}l_1^+)(l_2^-\bar\nu_2^
{\vphantom{+}}) e^-$, and their scattering amplitudes are:  
\begin{subequations}
\begin{multline}
-2s\Gamma_{ll Z}\frac{1}{t_1-M_Z^2}\\
\left[\frac{\Gamma_{ZW(W\to \nu_1^{\vphantom{+}}l_1^+)}(q_1,-q,k_1,k_2)}{M_1^2-M_W^2}\frac{1}{\that-M_W^2}\frac{\Gamma_{WZ(W\to l_2^-\bar\nu_2^{\vphantom{+}})}(q,q_2,k_3,k_4)}{M_2^2-M_W^2}\right]\frac{1}{t_2-M_Z^2}\Gamma_{ll Z},
\label{example}
\end{multline}
\begin{multline}
-2s\Gamma_{ll \gamma}\frac{1}{t_1}
\left[\frac{\Gamma_{\gamma W(W\to \nu_1^{\vphantom{+}}l_1^+)}(q_1,-q,k_1,k_2)}{M_1^2-M_W^2}\frac{1}{\that-M_W^2}\frac{\Gamma_{W\gamma(W\to l_2^-\bar\nu_2^{\vphantom{+}})}(q,q_2,k_3,k_4)}{M_2^2-M_W^2}\right]\frac{1}{t_2}\Gamma_{ll\gamma}.
\end{multline}
\end{subequations}
Finally, the subprocesses $WZ\to ZW$, $WZ\to \gamma W$, $W\gamma\to ZW$ and $W\gamma\to \gamma W$ are contained in 
the $2\to 6$ scattering processes $e^+e^-$ $\to$ 
$\bar\nu_e (l_1^-l_1^+)(\nu_2^{\vphantom{+}} l_2^+) e^-$. The corresponding 
scattering amplitudes can be constructed from the vertices listed above.
Note that vector boson scattering processes which are made up of just 
$Z$'s and $\gamma$'s, such as $\gamma\gamma\to ZZ$, do not exist on the tree 
level but require at least one loop.

As we have said before, inside these physical $2 \to 6$ inelastic processes 
the $2 \to 2$ vector-vector scattering processes appear as off-shell 
subprocesses, and they contain both longitudinal and transverse polarizations 
of the vector bosons. As an example, let us take the $2 \to 6$ process in 
eq.(\ref{eq2n4n6}), and consider the effective production vertex 
in (\ref{eqgammahat}). As explicitly shown in \cite{Ba}, the produced 
$W$ vector boson, in the overall center of mass system, can have both 
transverse or longitudinal polarization. Transforming to the center of 
mass system of the produced pair of $W$ bosons, we still have all three 
polarization states of the two $W$ bosons. Moving on to the particle 
poles in the $t_1$ and $t_2$ channels and making use of helicity conservation,
we conclude that we are dealing with a superposition of transverse and 
longitudinally polarized vector bosons, both in the `incoming' 
$t_1$ and $t_2$ channels and in the `outgoing' $M_1$ and $M_2$ channels.   

\label{polarization}
A separation of the different polarization starts from the angular 
distribution of the two fermion decay of the `outgoing' vector boson.     
Let us again consider the $WW \to WW$ process contained in eq.(\ref{eq2n4n6}).
Moving into the rest frame of one of the `outgoing' $W$ bosons, and 
denoting the decay angles of the two decay products by $\theta_i$ and 
$\pi -\theta_i$, the decay amplitude of an \ons\ transversal 
polarized $W$ behaves like $1\pm\cos \theta_i$, 
while for a logitudinal polarized $W$ it is proportional to  $\sin\theta_i$ 
\cite{Hagiwara:1986vm,Duncan:1985vj}. Hence, by a suitable 
projection we can separate the different outgoing polarization states. 
An analogous statement holds also for `outgoing' $Z$ bosons.  
By further making use of helicity conservation at the effective production 
vertex, this separation of polarization extends also to the `incoming' vector 
bosons of the quasielastic subprocess. This brief argument illustrates 
that, by analysing the angular distribution of the decay products of the
`outgoing' vector boson and by moving onto the particle poles in the 
$t_i$ and $M_i$ channels, it is possible to regain the usual on-shell 
vector-vector scattering amplitudes and, in particular, to isolate 
longitudinal vector-vector scattering.

\begin{figure}
\begin{center}\includegraphics{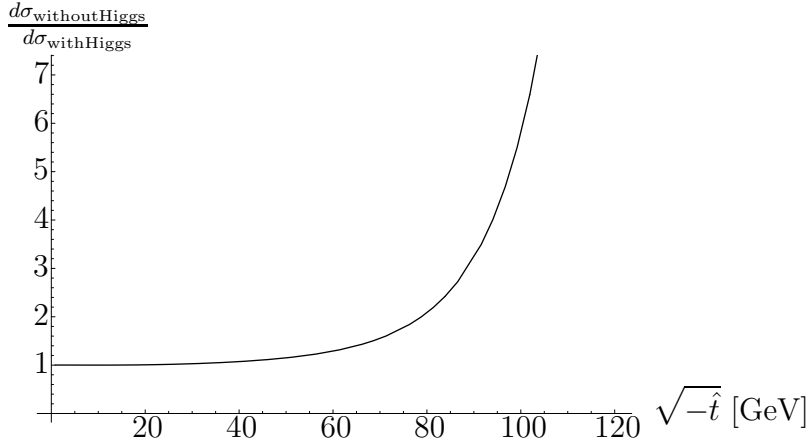}\end{center}
\caption{The $\that$ dependence of the ratio of $\frac{d\sigma }{d\that}$.} 
\label{plotratio}
\end{figure}

Let us finally return to the specific case of $WW$ scattering.
In section 3, as a first application of our formulae, we will be interested 
in the dependence on the Higgs mass at high energies. In the absence of the 
Higgs contribution, \eqref{eq2n4n6} is replaced by 
\begin{align}
T^{(s, {\rm without Higgs})}_{2 \to 4\to 6} =& -2s \Gamma_{l\nu W} \frac{1}{t_1-M_W^2}\nonumber\\
& \Bigg[ 
\frac{\Gamma_{WZ(W\to \nu_1^{\phantom{+}}l_1^+)}(q_1,-q,k_1,k_2)}{M_1^2-M_W^2}\;
\frac{1}{\hat{t}-M_Z^2}\;
\frac{\Gamma_{ZW(W\to l_2^-\bar\nu_2^{\phantom{-}})}(q,q_2,k_3,k_4)}{M_2^2-M_W^2} \nonumber\\ 
& +
\frac{\Gamma_{W \gamma (W\to \nu_1^{\phantom{+}}l_1^+)}(q_1,-q,k_1,k_2)}{M_1^2-M_W^2}\;
\frac{1}{\hat{t}} \;
\frac{\Gamma_{\gamma W(W\to l_2^-\bar\nu_2^{\phantom{-}})}(q,q_2,k_3,k_4)}{M_2^2-M_W^2}\nonumber\\
&-g_w^2M_W^2\frac{\bar{u}_{\nu_1} \gamma_\mu d v_{l_1^+}}{M_1^2-M_W^2}\;
\frac{\bar{u}_{l_2^-} \gamma^\mu d v_{\bar\nu_2}}{M_2^2-M_W^2}\Bigg]\frac{1}{t_2-M_W^2} \Gamma_{l\nu W}.
\label{eq2n4n6nohiggs}
\end{align}
Comparing \eqref{eq2n4n6} with \eqref{eq2n4n6nohiggs} one easily sees 
that the presence of the Higgs particle affects the dependence upon  
$\that$: in the large $\that$ region the renormalizibility of the 
electroweak theory requires the amplitude to vanish, and it is the Higgs
particle which cancels the constant term in square bracket term in 
\eqref{eq2n4n6nohiggs}. We illustrate this effect of the Higgs mass 
numerically in Fig.\ref{plotratio}: the ratio of $\frac{d\sigma}{d\that}$ for the theory without Higgs vs. the theory with Higgs becomes large when $\that$ exceeds $M_W^2$.
For this plot we choose $M_H=115$GeV, and the phase space integration 
covers the region $t_1,t_2 \in [-M_W^2,-.01 M_W^2]$, 
$\delta_{1} = \frac{M_{1}^2-M_W^2}{M_W^2}$, 
$\delta_{2} = \frac{M_{2}^2-M_W^2}{M_W^2}$ $\in [.025, .1]$, 
$\sqrt{\shat} \in [400{\rm GeV},500{\rm GeV}]$, $s_{01234}\in 
[400{\rm GeV},750{\rm GeV}]$  where $s_{01234}=(p_A'+k_{12}+k_{34})^2$.

For a heavy Higgs it is instructive to consider also next-to-leading 
contributions to \eqref{eq2n4n6} which are suppressed by a factor 
$\frac{M_H^2}{\shat}$. It is these terms which will lead to a conflict with 
the unitarity bounds once the Higgs become too heavy. For the case of 
$WW \to WW$ scattering we find: 
\begin{equation}
T^{(H)}_{2\to 4\to 6} = 
-2s\Gamma_{l\nu W} \frac{1}{t_1-M_W^2}\Bigg[\frac{g_w^2M_W^2M_H^2}{\shat}\;
\frac{\bar{u}_{\nu_1} \gamma_\mu d v_{l_1^+}}{M_1^2-M_W^2}\;
\frac{\bar{u}_{l_2^-} \gamma^\mu d v_{\bar\nu_2}}{M_2^2-M_W^2}\Bigg]\frac{1}{t_2-M_W^2}\Gamma_{l\nu W}.
\label{eq2n4n6higgs}
\end{equation}
The analogous Higgs mass dependent next-to-leading expressions for two other 
processes, namely for $WW\to ZZ$ scattering inside the $2\to6$ process 
\eqref{example2}:
\begin{equation}
-2s \Gamma_{l\nu W} \frac{1}{t_1-M_W^2}
\Bigg[\frac{g_w^2M_W^2M_H^2}{\shat}\;\frac{\bar{u}_{l_1^-} d_\mu  v_{l_1^+}}{M_1^2-M_Z^2}\;\frac{\bar{u}_{l_2^-} d^\mu  v_{l_2^+}}{M_2^2-M_Z^2}\Bigg]\frac{1}{t_2-M_W^2} \Gamma_{l\nu W}\label{eqT2n4n6WWZZ}
\end{equation}
and for $ZZ\to WW$ scattering inside \eqref{example} are:
\begin{equation}
-2s \Gamma_{ll Z} \frac{1}{t_1-M_Z^2}
\Bigg[\frac{g_w^2M_W^2M_H^2}{\shat}\;\frac{\bar{u}_{\nu_1} d_\mu  v_{l_1^+}}{M_1^2-M_W^2}\;\frac{\bar{u}_{l_2^-} d^\mu  v_{\bar\nu_2}}{M_2^2-M_W^2}\Bigg]\frac{1}{t_2-M_Z^2} \Gamma_{ll Z}.\label{eqT2n4n6ZZWW}
\end{equation}

\section{Unitarity Bounds}
\label{unitarity}
As a first application of our analytic expressions, we reconsider the
derivation of unitarity bounds on the Higgs mass; in contrast to 
earlier discussions in the literature which did not include the photon pole 
we will examine the influence of the photon pole. As we will discuss in 
detail, our discussion will be much less rigorous than the traditional 
arguments based upon on-shell scattering.    

We start by showing, for on-shell vector-vector scattering, 
how the usual investigation of the fixed-angle limit can be 
carried out also in the Regge limit. Unitarity of the $2 \to 2$ scattering 
process is illustrated in Fig. \ref{figelasticunitarity}. The signs within the bubbles indicate 
on which side of the branch cut in the complex $s$ plane the amplitudes 
has to be evaluated. A ``$+$'' marks the value above the cut, 
$s\underline{+i\epsilon}$,  a ``$-$'' below the cut, 
$s\underline{-i\epsilon}$. Projecting on the partial waves one arrives 
at the inequality: 
     
\begin{figure}
\begin{center}\includegraphics{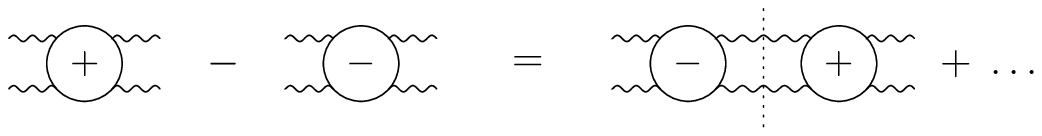}\end{center}
\caption{unitarity of $2 \to 2$ scattering}
\label{figelasticunitarity}
\end{figure}
\begin{align}
\imag T_l^{(el)} =& \left|T_l^{(el)}\right|^2+\sum_n \left|T_l^{(inel)}\right|^2\label{eqpraeunitaritypartialwave}\\
\Rightarrow\quad \left|T_l^{(el)}\right|\ge \left|\imag T_l^{(el)}\right| \ge & \left|T_l^{(el)}\right|^2\\
\Rightarrow\quad \left|T_l^{(el)}\right| \le & 1.
\label{equnitaritypartialwave}
\end{align}
Lee, Quigg and Thacker \cite{Lee:1977eg} have used this inequality in order 
to derive, from the elastic scattering of longitudinal polarized $W$ bosons,
an upper bound on the Higgs mass. The $l=0$ partial wave is obtained from 
the fixed-angle limit of the elastic scattering amplitude, and in the 
absence of a Higgs particle it has the form:  
\begin{equation}
T_0 ^{(\rm without Higgs)}(W^+_LW^-_L) = \frac{g_w^2}{128\pi} \frac{s}{M_W^2}+\orders{0} .\label{eqT0withoutH}
\end{equation}
For energies of the order 
$\sqrt{s}\simeq 16\sqrt{\frac{\pi}{2}}\frac{M_W}{g_w} \approx 2.4 \text{TeV}$
the inequality (\ref{equnitaritypartialwave}) is violated, and the $W$'s 
start to interact strongly \cite{Chanowitz:1998wi}. Once the Higgs is 
included, the term proportional to $s$ cancels. But even though the $l=0$ 
partial wave is no longer growing with $s$, this does not automatically 
imply that it is smaller than 1. Lee, Quigg and Thacker calculated the 
matrix element in the \fal\ using the equivalence theorem 
\cite{Cornwall:1974km,Veltman:1990ud} and found that the 
0th partial wave is proportional to the Higgs mass squared:
\begin{equation}
T_0 ^{(\rm with Higgs)}(W^+_LW^-_L) = \frac{g_w^2}{32\pi} \frac{M_H^2}{M_W^2}+\orders{-1}. \label{eqT0withH}
\end{equation}
The inequality (\ref{equnitaritypartialwave}) leads to the condition
$M_H < 8\sqrt{\frac{\pi}{2}}\frac{M_W}{g_w} \approx 1.2 \text{TeV}$. This 
bound can be made more stringent if the scattering channels $ZZ$ and $HH$ are 
included. One finds $\Rightarrow M_H < 8\sqrt{\frac{\pi}{3}}\frac{M_W}{g_w} 
\approx 1.0 \text{TeV}$. In this discussion the photon exchange had not 
been included. As function of 
the scattering angle, the photon propagator goes like 
$\sim\frac{1}{1-\cosq}$, and the partial wave projection is not defined, 
unless the region $\theta \sim 0$ is excluded. 

It is not difficult to derive the bound on the Higgs particle also from 
the Regge limit: $s \to \infty$, $t$ fixed (i.e. $\theta \to 0$).  
Instead of the partial wave we define the Fourier transform with respect to 
impact parameter $\bb$:
\begin{align}
\widetilde{T}(s,b) =& \frac{1}{16\pi^2s}\int d^2 \bq e^{-i\vec{q}_\perp\vec{b}}T\left(s,t\right).\label{eqtvonb}
\end{align}
Within this representation the unitarity relation can be simplified in a 
way similar to the partial wave decomposition:
\begin{align}
\imag \widetilde{T}(s,b)^{(el)} =& \left|\widetilde{T}(s,b)^{(el)}\right|^2+\sum_n \left|\widetilde{T}(s,b)^{(inel)}\right|^2\\
\Rightarrow\quad \left|\widetilde{T}(s,b)^{(el)}\right| \le  & 1. \label{equnitarityimpact}
\end{align}
In the Regge limit the matrix element for $W^+_LW^-_L$ scattering without a 
Higgs contribution reads:
\begin{align}
T ^{(\rm without Higgs)}=& -2 g_w^2 s 
\left[\frac{c_w^2}{t-M_Z^2}\left(\frac{2 M_W^2-M_Z^2}{2M_W^2}\right)^2+
\frac{s_w^2}{t} \right] \nonumber\\
&  -\frac{g_w^2s}{2M_W^4}\left[-\frac{M_W^2}{2}\right]+\order{s^0}. \label{TrgwithoutHiggs}
\end{align}
Ignoring, for the time being, the photon pole, one sees that the violation 
of unitarity now arises due to the the large-$t$ behavior: since 
in the $b$-transform the $\bq$-integral can not be extended to infinity, 
we have to remember that $-t_{\rm max}=s-4M_W^2$. The $b$ transform 
now grows with $s$ and violates the unitarity bound. 
Once the Higgs is included the scattering amplitude takes the form    
\begin{align}
T ^{(\rm with Higgs)} =& -2 g_w^2 s \left[\frac{c_w^2}{t-M_Z^2}
\left(\frac{2M_W^2-M_Z^2}{2M_W^2} \right)^2+\frac{s_w^2}{t} \right] \nonumber\\
&  -\frac{g_w^2}{2M_W^4}\left[\frac{M_H^2 M_W^2}{2}\right]+\order{s^0 M_H^0}. 
\label{TrgwithHiggs}
\end{align}
The term proportional to $s$ now vanishes as $\that$ becomes large.
As a consequence, the $\bq$-integral converges, with the exception of the 
point $\bb =0$: at this point the $b$-transform grows as $\ln s/M_W^2$, but 
this contribution becomes significant only at energies much higher than 
the TeV-region and, in our discussion, can safely be neglected. 
For a heavy Higgs, however, the second term in (\ref{TrgwithHiggs}) 
becomes significant: 
the contribution from $-t \sim s-4 M_W^2$  exceeds unity unless the Higgs 
mass stays below the upper limit which is the exactly the same 
(cf. eq.(\ref{eqT0withH})) one as the one obtained by Lee, Quigg, and Thacker. 
As in the fixed angle limit, the photon pole can not be 
included, since the $b$ integral diverges at $\bq=0$. 
        
\begin{figure}[htbp]
\begin{center}\includegraphics{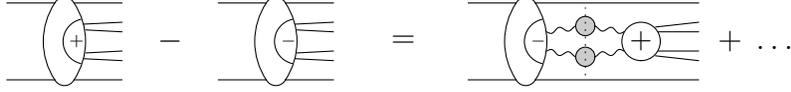}\end{center}
  \caption{unitarity of $2\to 6$ scattering}
  \label{figunitarityinelastic}
\end{figure}

Turning now to the off-shell scattering of vector bosons,
we immediateley see that it is the $\thatmin$ in eq.(\ref{tmin}) 
which excludes the dangerous photon pole. The question we are interested in 
is the influence of the photon exchange on the unitarity 
bound of the Higgs mass. In order to derive a unitarity 
bound we have to consider the singularity structure of six-point amplitude 
$T_{2 \to 6}$ in the energy variable $\shat$ which plays the role of the 
energy of the $WW$ scattering subprocess. The corresponding unitarity 
integral is illustrated in Fig.\ref{figunitarityinelastic}.         

Again, the '$\pm$' signs denote on which side of the branch cut the amplitudes have to be 
evaluated. The intermediate states on the rhs of the unitarity equation 
consist of stable particles. Since, in our off-shell discussion, 
the $W$ boson has to be treated as unstable ~\cite{Veltman:1963th}, 
the sum over intermediate states  
has to extend over the stable decay products of the $W$ boson, which, 
to lowest order, consists of a pair of leptons, $(l\nu)$. In our figure, 
we have marked the cut through the self energy by the a blob with a 
vertical cutting line. The $W$ propagators on the rhs and on the lhs of this 
cut blob 
contain the self energies, i.e. the poles are slightly off the real axis.
While in the unitarity integral the decay 
products of the $W$ boson are kept on-shell, their momenta are integrated over.
As a result, the sqares of the four momentum of the $W$ bosons, 
$k_{12}^2=M_1^2$ and $k_{34}^2=M_2^2$, also vary.
Moreover, they are real-valued and thus never reach the $W$ pole in the  
complex plane. 

All this discussion applies to a situation where all orders 
in perturbation theory are taken into account. Our discussion is carried out 
in lowest order perturbation theory: in this 
approximation the $W$ pole lies on the real axis, and intermediate state 
on the rhs in Fig. \ref{figunitarityinelastic} reduces to the $W$ boson. Nevertheless, thanks to the 
off-shellness of the `incoming' and `outgoing' $W$ bosons, 
$\thatmin$ stays below zero, and the integration over the intermediate 
state momenta leads to logarithm of $\thatmin$. In order to be as close 
as possible to the realistic situation, we will chose $k_{12}^2=M_1^2$ and 
$k_{34}^2=M_2^2$ to differ from $M_W^2$ by a characteristic size 
$\Gamma_W M_W$.

In contrast to the $2 \to 2$ on-shell scattering it is not possible to 
derive, on general grounds, a rigorous inequality analogous to 
(\ref{equnitaritypartialwave}) or (\ref{equnitarityimpact}). 
Since the scattering amplitude $T_{2 \to 6}$ 
depends upon several energy variables with nonvanishing discontinuities, 
the discontinuity in $\hat{s}$ does not coincide with the imaginary 
part. All we can say is that, in leading order in $s/M_W^2$ and in
$M_H^2/M_W^2$, the scattering amplitude of our process, $T_{2 \to 6}$, is 
real-valued; all energy discontunities are suppressed by one power in 
$g_W^2$. The imaginary part is a sum of energy discontinuities. 
Because of the linear independence of the different energy discontinuities, 
a large discontinuity in $\shat$ 
can not be compensated by a discontinuity in another energy variable: 
a too strong growth of the discontinuity in $\shat$, 
therefore, should certainly be viewed as indicative of violation of 
unitarity. 

Furthermore, the `inelastic' contribution on the rhs of the discontinuity  
equation (Fig.6) for $T_{2 \to 6}$ can no longer 
be written as a sum of squares. Its contribution to the unitary sum 
could, in principle, have the opposite sign compared to the elastic 
intermediate state; this, however, would require a rather dramatic difference 
between the on-shell $WW$ scattering amplitude and the off-shell situation.
When starting from the $W$-poles in the $t_1$, $t_2$, $M_{1}^2$, and $M_{2}^2$ 
channels, our $T_{2 \to 6}$ amplitude contains the on-shell 
$WW$ scattering amplitude which, in its own energy discontinuity, contains 
a positive inelastic contribution. Moving away from these poles, 
the inelastic contribution to the discontinuity in $\shat$ would have to 
decrease, pass through zero and become negative. This would be fairly 
strong variation.
 
Finally a remark on the polarization. As we have discussed before, the 
$2 \to 6$ amlitudes on the lhs 
of Fig.6 contains both longitudinal and transverse polarizations of the 
produced vector bosons. By a suitable analysis of the angular distribution 
of the fermion pair one could isolate the longitudinal part; in our subsequent 
estimate of the effect of the photon pole we will not do this. On the other 
hand, in the intermediate states on the rhs of the unitarity equation 
in Fig.6 , we will restrict ourselves to longitudinal polarized $W$ bosons 
only. We expect that, as far as the unitarity bound for large Higgs masses is 
concerned, the neglect of the transverse polarized intermediate states 
which are not affected by the Higgs mass will be a small effect.
 
Summing up this discussion, it seems plausible (but not proven) that we still can use the inequality  
\begin{equation}
\left|\widetilde{T}_{2\to 4\to 6}(\bhat)\right| \ge
\left|\widetilde{T}_{2\to 4}(\bhat)\widetilde{T}_{2\to 2\to 4}(\bhat)\right|,
\label{equnitaitaetofs}
\end{equation}
where $T_{2\to 4}$ denotes the process $e^+e^-$ $\to$ 
$\bar\nu_e W^+_L W^-_L \nu_e$, and $T_{2\to 2\to 4}$ stands for the process 
$W^+_L W^-_L$ $\to$ $\nu_1l_1^+ l_2^-\bar\nu_2$. The `impact parameter' 
vector $\mbf{\bhat}$ belongs to the exchange in the $WW$-subsystem. 
In the following we shall examine and numerically estimate the role of the 
photon pole in \eqref{equnitaitaetofs}.

\begin{figure}[htbp]
\begin{center}\includegraphics{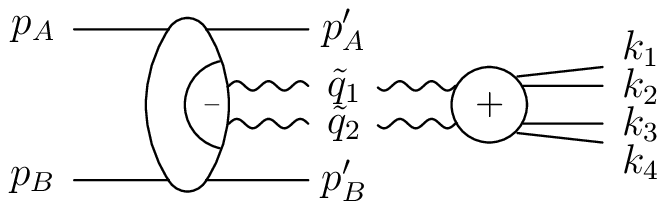}\end{center}
  \caption{kinematics of $T_{2\to 4}$ and $T_{2\to 2\to 4}$}
  \label{figurekinematics}
\end{figure}

The kinematic variables of the $2\to 4$-process 
$e^+(p_A)e^-(p_B)$ $\to$ $\bar\nu_e(p'_A)W_L^+(\tilde{q}_{1})W_L^-
(\tilde{q}_{2})\nu_e(p'_B)$ are the same as those of the $2\to 6$-process, 
except for the outgoing states of the subsystem which are shown in figure 
\ref{figurekinematics}. The momentum transfer $\that'$ in the $WW$ subsystem 
is given by $\that ' \equiv {q'}^2 \equiv (q_1-\tilde{q}_{1})^2$.
The momentum transfer $\that''$ in the $2\to 4$-process
$W_L^+(\tilde{q}_1)W_L^-(\tilde{q}_2)$ $\to$ $\nu_1(k_1)l_1^+(k_2) 
l_2^-(k_3)\bar\nu_2(k_4)$ is given by $\that'' \equiv {\tilde{q}}^2 
\equiv (\tilde{q}_1-(k_1+k_2))^2$.
The scattering amplitudes which enter the lhs of 
\eqref{equnitaitaetofs} can be extracted from the corresponding 
$2\to 6$ process \eqref{eq2n4n6} which are described in section 2.

Let us make a few remarks 
on the kinematics. First we notice that, on both sides of 
eq.(\ref{equnitaitaetofs}), we have the denominators     
$(t_{1}-M_W^2)$, $(t_{2}-M_W^2)$,  and $(M_{1}^2-M_W^2)$, $(M_{1}^2-M_W^2)$.
It is convenient to define new functions 
$\widetilde{\mathcal{T}}(\bhat)$: 
\begin{align}
\widetilde{\mathcal{T}}_{2\to 4\to 6} =& \frac{\hat{s}}{s}(t_1-M_W^2)(t_2-M_W^2)(M_1^2-M_W^2)(M_2^2-M_W^2)\widetilde{T}_{2\to 4\to 6}\\
\widetilde{\mathcal{T}}_{2\to 4} =& \frac{\hat{s}}{s}(t_1-M_W^2)(t_2-M_W^2)\widetilde{T}_{2\to 4}\\
\widetilde{\mathcal{T}}_{2\to 2\to 4} =& (M_1^2-M_W^2)(M_2^2-M_W^2)\widetilde{T}_{2\to 2\to 4},
\end{align}
where, for brevity, we have dropped the dependence upon $b$.  
Inserting these functions into (\ref{equnitaitaetofs}) we obtain:
\begin{equation}
\quad|\widetilde{\mathcal{T}}_{2\to 4\to 6}| \ge |\widetilde{\mathcal{T}}_{2\to 4}\widetilde{\mathcal{T}}_{2\to 2\to 4}|.
\label{equnitaritaetofsshort}
\end{equation}

Following the discussion of the on-shell elastic scattering amplitude, we 
begin with the case where the Higgs particle is absent, and we ignore the 
photon exchange. We consider the point $\bb=0$ which is most sensitive 
to the large-$t$ behavior. 
On the lhs of (\ref{equnitaritaetofsshort}) we find, from the constant terms 
in (\ref{eq2n4n6nohiggs}), that the large-$\that$ 
region leads to a growth proportional to $\shat/M_W^2$. Similarly, on the rhs 
both the first and the second factor grow as $\shat/M_W^2$ and violates the 
inequality which follows from unitarity. 
Once the Higgs particle is included the growth proportional to 
$\shat/M_W^2$ disappears on both sides of (\ref{equnitaritaetofsshort}). 
The Fourier transform to $b$ space has to be done numerically; 
however it is not difficult to understand the general pattern by 
concentrating on those terms which become large as both $M_H$ and $\hat{s}$ 
grow. 
To begin with the lhs, the term proportional to 
$M_H^2$ leads to a contribution which is constant in $\shat$; 
its coefficient follows directly from (\ref{eq2n4n6higgs}).
Next, the $Z$ exchange in (\ref{eq2n4n6}) 
leads to a term which grows proportional to $\ln M_Z^2/\shat$.
Finally, the photon exchange yields an additional term proportional to $\ln \thatmin/\shat$ 
 as well as terms proportional to $\ln \thatmin/M_W^2$ where $\thatmin$ is given in (\ref{tmin}).
The coefficient of the terms proportional to $\ln \thatmin/M_W^2$ has a somewhat complicated form and has to be evaluated numerically.
On the rhs we find, for each of the two scattering ampltitudes, 
four terms of the same kind. As a result, 
our inequality (\ref{equnitaritaetofsshort}) takes the 
form:
\begin{multline}
       \frac{g_w^6}{32\pi}\left(M_H^2M_W^2 + c_w^2(2M_W^2-M_Z^2)^2 \ln \frac{M_Z^2}{\shat} + 4s_w^2M_W^4 \ln \frac{\thatmin}{\shat}+C_1 \ln\frac{\thatmin}{M_W^2}\right) \ge \\
 \frac{g_w^4}{16\pi}\left(M_H^2  + c_w^2\frac{(2M_W^2-M_Z^2)^2}{M_W^2} \ln \frac{M_Z^2}{\shat} + 4 s_w^2M_W^2 \ln \frac{\thatmin'}{\shat}+C_2\ln\frac{\thatmin}{M_W^2}\right) \\
\cdot\frac{g_w^4}{64\pi}\left(M_H^2 + c_w^2\frac{(2M_W^2-M_Z^2)^2}{M_W^2} \ln \frac{M_Z^2}{\shat} + 4 s_w^2M_W^2 \ln \frac{\thatmin''}{\shat}+C_3\ln\frac{\thatmin}{M_W^2}\right) ,\label{analyticinequality}
\end{multline}
where 
\begin{equation}
\that'_{min} = \frac{(t_1 - M_W^2)(t_2-M_W^2)}{\shat},\;
\that''_{min} = \frac{(M_1^2 - M_W^2)(M_2^2-M_W^2)}{\shat}.
\label{result}
\end{equation}
The three positive coefficients $C_i$ have to be computed numerically:
In each of the three factors, the main contribution comes from the 
first term proportional to $M_H^2$: if the other terms would be ignored, 
the Higgs bound would be the same as in the on-shell case. The terms 
$\sim \ln \thatmin$, are due to the photon exchange; for small $\thatmin$, 
they have the opposite sign and they tend to weaken the bound on the Higgs 
mass. 

\begin{figure}
\begin{center}\includegraphics{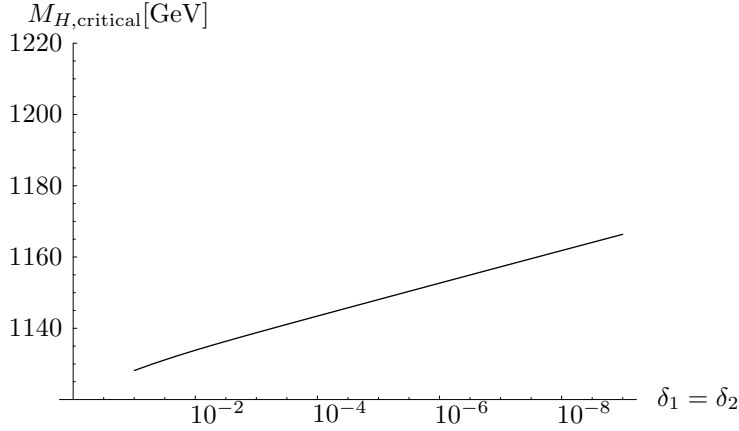}\end{center}
\caption{The critical Higgs mass as a function of 
$\delta_{1}=\delta_{2}$ at $t_{1}=t_{2}= -M_W^2$, $\sqrt{\shat}=10^3$GeV.}
\label{figuredelta}
\end{figure}

Further results are obtained from a numerical analysis.
By setting $\theta_i=\frac{\pi}{2}$ 
(see discussion on page \pageref{polarization}) we make the longitudinal 
component of the outgoing \ofs\ $W$'s maximal, i.e. we strengthen the 
sensitivity to the Higgs mass. 
Compared to the critical value of the Higgs mass, which is obtained 
in the \ons\ analysis, $1234$GeV, the bound in our off-shell analysis 
slightly increases up to the range between $1247$GeV and $1252$GeV (depending 
upon the values for $|t_1|$, $|t_2|$, and $|\delta_1|$, $|\delta_2|$ which are varied over the intervals described at the end of section 2).
In the inequality (\ref{analyticinequality}), the contribution 
which is most sensitive to the photon exchange comes from the last 
term on the rhs $\propto \ln\frac{\thatmin''}{\shat}$ with $\thatmin''=\frac{\delta_1 \delta_2 
M_W^4}{\shat}$. For the considered ranges of $\delta_{1}$, $\delta_{2}$ and 
$\shat$ the parameter $\thatmin''$ varies between $.1{\rm Gev}^2$ and 
$2.6{\rm GeV}^2$. The rather weak dependence of the critical Higgs mass 
on $\delta_{1}$ and $\delta_{2}$ is illustrated in Fig. \ref{figuredelta}.   

We end this section by including into our numerical analysis the 
intermediate states of other weak vector bosons and of the Higgs boson. 
In the on-shell analysis, the bound on the Higgs mass becomes stronger 
once these vector and Higgs states have been included, not only in the
intermediate states but also in the initial and final states. In our 
off-shell analysis we will not go quite as far, since different outgoing 
vector bosons also would affect the decays into fermion pairs. Instead, we 
stick to incoming and outgoing $W$'s, and we only include additional 
$Z$'s and Higgs particles in the intermediate states:

\begin{equation}
\left| \widetilde{\mathcal{T}}_{2\to 4\to 6}\right| \ge 
\left|\widetilde{\mathcal{T}}_{2\to 4}^{(WW)}\widetilde{\mathcal{T}}_{2\to 2\to 4}^{(WW)} + 
\frac{1}{2}\widetilde{\mathcal{T}}_{2\to 4}^{(ZZ)}\widetilde{\mathcal{T}}_{2\to 2\to 4}^{(ZZ)} +
\frac{1}{2}\widetilde{\mathcal{T}}_{2\to 4}^{(HH)}\widetilde{\mathcal{T}}_{2\to 2\to 4}^{(HH)}\right|.
\label{equnitaritaetofsverbessert}
\end{equation}

Here the upper indices refer to the particles in the (longitudinal) 
intemediate states, e.g.     
$\widetilde{\mathcal{T}}_{2\to 4}^{(ZZ)}$ belonges to the process 
$e^+e^-$ $\to$ $\bar\nu_e Z_L Z_L \nu_e$, whereas 
$\widetilde{\mathcal{T}}_{2\to 2\to 4}^{(ZZ)}$ is part of the process 
$Z_L Z_L$ $\to$ $\nu_1 l_1^+ l_2^- \bar\nu_2$. 
The  kinematic variables for the process  $e^+(p_A)e^-(p_B)$ $\to$ 
$\bar\nu_e(p'_A)Z_L(\tilde{q}_{1})Z_L(\tilde{q}_{2})\nu_e(p'_B)$ are the 
same as those of the $2\to 6$-process, except for the outgoing states of 
the subsystem (momentum transfer $\that ' \equiv {q'}^2 \equiv 
(q_1-\tilde{q}_{1})^2$). The matrix elements can be extracted from the 
associated $2\to 6$-amplitude \eqref{example2} in the same way as in 
the $WW\to WW$ case. In the same we derive from \eqref{example}
the matrix element for the process $Z_L(\tilde{q}_{1})Z_L(\tilde{q}_{2})$ 
$\to$ $\nu_1(k_1)l_1^+(k_2) l_2^-(k_3)\bar\nu_2(k_4)$.

The matrix element for the process  $e^+(p_A)e^-(p_B)$ $\to$ 
$\bar\nu_e(p'_A)H(\tilde{q}_{1})H(\tilde{q}_{2})\nu_e(p'_B)$ reads 
(momentum transfer $\that ' \equiv {q'}^2 \equiv (q_1-\tilde{q}_{1})^2$):
\begin{equation}
T_{2\to 4}^{(HH)} =
2s \Gamma_{l\nu W}\frac{1}{t_1-M_W^2}\Bigg[g_wM_W\;\frac{1}{(\that'-M_W^2)}\;g_wM_W+\frac{g_w^2M_H^2}{2\shat}\Bigg]\frac{1}{t_2-M_W^2}\Gamma_{l\nu W}.
\end{equation}
The matrix element for the process $H(\tilde{q}_1)H(\tilde{q}_2)$ $\to$ 
$\nu_1(k_1)l_1^+(k_2) l_2^-(k_3)\bar\nu_2(k_4)$ reads (momentum transfer 
$\that'' \equiv {\tilde{q}}^2 \equiv (\tilde{q}_1-(k_1+k_2))^2$):
\begin{equation}
T_{2\to 2\to 4}^{(HH)} =
2\shat\Bigg[\frac{g_wM_W\bar{u}_{\nu_1} \gamma_\mu  d  v_{l_1^+}}{M_1^2-M_W^2}\;\frac{1}{\that ''-M_W^2}\;\frac{g_wM_W\bar{u}_{l_2^-} d^\mu  v_{\bar\nu_2}}{M_2^2-M_W^2}+\frac{g_w^2M_H^2}{2\shat}\;\frac{\bar{u}_{\nu_1} \gamma_\mu d v_{l_1^+}}{M_1^2-M_W^2}\;\frac{\bar{u}_{l_2^-} d^\mu  v_{\bar\nu_2}}{M_2^2-M_W^2}\Bigg].
\end{equation}
In contrast to the coupling of $Z$ exchange to incoming 
$W$ bosons, $\Gamma_{WZ(W\to \nu_1l_1^+)}$ 
in \eqref{eqVVffVertex}, the analogous vertex for incoming Higgs bosons  are 
very simple because in all non resonant diagrams the Higgs couple to fermions. 
These couplings are proportional to the fermion masses, and they can be 
neglected in comparison with the gauge boson masses.

With the same kinematical parameters used before, in this extended analysis  
the critical Higgs mass lies between $1119{\rm GeV}$ and $1126{\rm GeV}$,
compared to the critical Higgs mass  $1104{\rm GeV}$ in the analogous 
on-shell analysis. The dependence upon $\shat$ and $\delta_{1}$, $\delta_{1}$
is very much the same as in the previous single-channel analysis. 

\section{Conclusions}

We have calulated analytic expressions for the matrix elements of 
off-shell scattering of weak vector bosons
in the limit of large center of mass energies. Our \ofs\ scattering 
amplitudes are derived from the process $e^+e^-$ 
$\to$ $\bar\nu_e\nu_1l_1^+ l_2^-\bar\nu_2\nu_e$ 
in the multi-\rgl . The expressions have a simple factorized form and can be 
used for further theoretical studies of $W^+W^-$ scattering. 
We also have calculated the related matrix elements for vector boson 
scattering processes composed of the channels $WW$, $ZZ$, $\gamma\gamma$, 
$WZ$, $W\gamma$, $Z\gamma$.

As an application, we have re-examinded the derivation of bounds on the Higgs 
mass from unitarity. In contrast to on-shell scattering, our 
off-shell analysis allows to include the photon exchange. We have given 
plausibility arguments which have lead us to an inequality, 
derived from unitarity. Based upon this inequality we have examined 
the influence of the photon exchange on the upper bound of the Higgs mass.  
As a result we have shown that the photon pole tends to weaken the 
bound on the Higgs mass. We have demonstrated numerically 
that the shift of the mass bound due to the photon is small compared 
to the general uncertainty of the bound derived from lowest order 
unitarity contribution.


\begin{thebibliography}{10}

\bibitem{Lee:1977eg}
B. W. Lee, C. Quigg and H. B. Thacker,
{\em Phys. Rev.}, {\bf D 16} (1977) 1519.

\bibitem{Cornwall:1974km}
J. M. Cornwall, D. N. Levin and G. Tiktopoulos,
{\em Phys. Rev.}, {\bf D 10} (1974) 1145.

\bibitem{Dawson:1985gx}
S. Dawson,
{\em Nucl. Phys.}, {\bf B 249} (1985) 42.

\bibitem{Veltman:1990ud}
H. G. J. Veltman,
{\em Phys. Rev.}, {\bf D 41} (1990) 2294.

\bibitem{Barger:1995cn}
V. D. Barger, K. Cheung, T. Han and R. J. N. Phillips,
{\em Phys. Rev.}, {\bf D 52} (1995) 3815 [arXiv:hep-ph/9501379].

\bibitem{Kuss:1996yv}
I. Kuss and H. Spiesberger,
{\em Phys. Rev.}, {\bf D 53} (1996) 6078 [arXiv:hep-ph/9507204].

\bibitem{Denner:1998kq}
A. Denner and T. Hahn,
{\em Nucl. Phys.}, {\bf B 525} (1998) 27 [arXiv:hep-ph/9711302].

\bibitem{Gangemi:1998vc}
F. Gangemi, G. Montagna, M. Moretti, O. Nicrosini and F. Piccinini,
{\em Eur. Phys.  J.}, {\bf C 9} (1999) 31 [arXiv:hep-ph/9811437].

\bibitem{Dittmaier:2002ap}
S. Dittmaier and M. Roth,
{\em Nucl. Phys.}, {\bf B 642} (2002) 307 [arXiv:hep-ph/0206070].

\bibitem{BFKL}
For equal masses see L. N. Lipatov, {\em Sov. J. Nucl. Phys.} {\bf 23} (1976) 338
[{\em Yad. Fiz} {\bf 23} (1976) 642].

\bibitem{Ba}
J. Bartels,
{\em Nucl.\ Phys.}, {\bf B 151} (1979) 293.

\bibitem{Hagiwara:1986vm}
K. Hagiwara, R. D. Peccei, D. Zeppenfeld and K. Hikasa,
{\em Nucl. Phys.},  {\bf B 282} (1987) 253.

\bibitem{Duncan:1985vj}
M. J. Duncan, G. L. Kane and W. W. Repko,
{\em Nucl. Phys.}, {\bf B 272} (1986) 517.

\bibitem{Chanowitz:1998wi}
M. S. Chanowitz,
(1998) [arXiv:hep-ph/9812215].

\bibitem{Veltman:1963th}
M. J. G. Veltman,
{\em Physica}, {\bf 29} (1963) 186.

\bibitem{Eden:1966a}
R. J. Eden, P. V. Landshoff, D. I. Olive and J. C. Polkinghorne,
{\em The analytic {$S$}-matrix}, 
Cambridge Univ. Pr, 1966.

\end{thebibliography}
\end{document}